\documentclass[a4paper,draft,reqno,12pt]{amsart}
\usepackage{amssymb,euscript,bbold}
\usepackage{array}
\renewcommand{\d}{\partial}
\newcommand{\half}{\tfrac12}
\newcommand{\reg}{\text{reg}}
\renewcommand{\gg}{{\ensuremath{\mathfrak g}}}
\newcommand{\gh}{{\ensuremath{\mathfrak h}}}
\newcommand{\gt}{{\ensuremath{\mathfrak t}}}
\newcommand{\sG}{{\ensuremath{\mathsf G}}}
\newcommand{\sT}{{\ensuremath{\mathsf T}}}
\newcommand{\sJ}{{\ensuremath{\mathsf J}}}
\newcommand{\sA}{{\ensuremath{\mathsf A}}}
\newcommand{\1}{{\ensuremath{\mathbb 1}}}
\newcommand{\C}{{\ensuremath{\mathbb C}}}
\newcommand{\R}{{\ensuremath{\mathbb R}}}
\newcommand{\Z}{{\ensuremath{\mathbb Z}}}
\DeclareMathOperator{\rank}{\mathrm{rank}}
\DeclareMathOperator{\ord}{\mathrm{ord}}
\DeclareMathOperator{\vol}{\mathrm{vol}}
\begin{document}

\title[]{D-branes in Kazama--Suzuki Models}
\author[]{Sonia Stanciu}
\address[]{\begin{flushright}Theoretical Physics Group\\
Blackett Laboratory\\
Imperial College\\
Prince Consort Road\\
London SW7 2BZ, UK\end{flushright}}
\email{s.stanciu@ic.ac.uk}
\thanks{Supported by a PPARC Postdoctoral Fellowship.}
\begin{abstract}
We investigate boundary states of D-branes wrapped around
supersymmetric cycles in Kazama--Suzuki models.  We show that the
geometry of the D-branes corresponds to a generalisation of calibrated
geometry.  We comment on the link with the geometry of the coset space
and discuss how T-duality maps between these boundary states.  
\end{abstract}
\maketitle
 
\section{Introduction}

The advent of D-branes forces us to reconsider the problem of
consistent string vacua in order to determine the manifolds which
allow consistent string propagation, both perturbative and
non-perturbative.  In particular, D-branes probe a new aspect of the
background geometry, namely the geometry of submanifolds.  For
instance, the requirement that the worldvolume theory of a D-brane
wrapping on a given submanifold of the compactification manifold be
supersymmetric determines the so-called {\em supersymmetric cycles\/},
which in the case of backgrounds without a $B$ field
\cite{BBS,OOY,BBMOOY} are examples of calibrated submanifolds in the
sense of \cite{HL,H}.  

D-branes have been extensively studied in the case when their
worldvolume is flat.  However the study of D-branes on curved spaces
(see for instance \cite{BBS,BSV,OOY,BBMOOY,KO,BT,BT2,AFOS}) has been
largely restricted to the case of Calabi--Yau manifolds and some
manifolds of exceptional holonomy.  Calabi--Yau spaces seem to 
be perfectly suited for the study of D-branes, as they provide us with
much of the necessary information for this kind of analysis; however
it is important to learn how much of it relies on exactly
the Calabi--Yau structure, and how much remains true in slightly
different or wider contexts. 

Group manifolds and coset models, in particular the $N{=}2$
Kazama--Suzuki models, are among the few known examples of explicit
solutions for exact string backgrounds, and as such they make prime
candidates for the study of D-branes in curved manifolds.  Here the
algebraic structure of the model imposes non-trivial consistency
requirements on the possible boundary conditions \cite{KO}.  However
the essential feature that distinguishes these backgrounds from the
ones based on Calabi-Yau manifolds is the presence, in this case, of a
nontrivial $B$ field.  One of the main purposes of this paper is to
investigate how the geometry of the supersymmetric cycles depends of
this fact.  Moreover in this case the geometric data is known exactly
and we do not have to restrict ourselves to working in the large
volume limit as is the case for Calabi--Yau manifolds.

In this paper we initiate a study of D-branes on coset manifolds by
analysing, in the context of type II string theory, D-branes on
Kazama--Suzuki models.  We therefore start in Section 2 by reviewing
the basic aspects of the Kazama--Suzuki models, in order to set the
notation and exhibit the superconformal structure of these models.  In
Section 3 we write down the consistency requirements that boundary
states of $N{=}2$ superconformal theories (SCFT) have to satisfy.
They determine two types of boundary conditions, much in the same way
as happens in the case of Calabi--Yau threefolds \cite{OOY}.  In
Section 4 we discuss a class of solutions for the case of the
Kazama--Suzuki models.  In Section 5 we examine the geometric
interpretation of these boundary states in terms of D-branes wrapped
around cycles in the coset space.  The corresponding submanifolds
saturate a bound which can be thought of as a geometric analogue of
the Bogomol'nyi bound.  The appropriate framework for describing these
submanifolds is given by a generalisation \cite{HL} of the concept of
calibration, which we discuss in Section 6.  In general, these cycles
are not volume-minimising, and we argue why this does not conflict
with the requirement that they be BPS.  As an application of these
results, in Section 7 we discuss the way abelian T--duality
transformations map these D-brane configurations.

\section{Kazama--Suzuki models}

Let us start by defining our background.  For this we consider a
compact simple Lie group $G$, and $H$ a subgroup with $\gh
\subset \gg$ the corresponding Lie algebras.  
By the Kazama--Suzuki model \cite{KS1,KS2,S} we mean the $N{=}2$
superconformal field theory constructed from the $N{=}1$ coset theory
$G/H$.  In order to define this model we need an $N{=}1$ affine
algebra $\widehat\gg_{N{=}1}$, with generators $(I_a,\Psi_a)$ and 
\begin{align}
I_a (z)I_b (w) &= \frac{k\Omega_{ab}}{(z-w)^2} + \frac{{f_{ab}}^c I_c
(w)}{z-w} + \reg~,\label{eq:affg1}\\
I_a (z)\Psi_b (w) &= \frac{{f_{ab}}^c \Psi_c (w)}{z-w} +
\reg~,\label{eq:affg2}\\
\Psi_a (z)\Psi_b (w) &= \frac{k\Omega_{ab}}{z-w} +
\reg~.\label{eq:affg3}
\end{align}
$\Omega$ denotes an invariant metric of $\gg$ and the parameter $k$ is
related to the level $x$ of the affine algebra by $k=x+g^*$, where
$g^*$ is the dual Coxeter number.  Further we need an $N{=}1$ affine
subalgebra $\widehat\gh$ of $\widehat\gg$, generated by the currents
$(I_i,\Psi_i)$, and whose OPEs are similar to \eqref{eq:affg1}
-\eqref{eq:affg3}, with metric $\Omega_{ij}$, the restriction
$\Omega|_{\gh}$ of $\Omega$ to $\gh\subset\gg$. 

For the coset theory to have $N{=}2$ superconformal symmetry,
$\gg/\gh$ must possess an $\gh$-invariant metric
$\Omega_{\alpha\beta}$, and a compatible, integrable, $\gh$-invariant
complex structure ${A_{\alpha}}^{\beta}$ (see \cite{FS1} for details).
Then the generators of the $N{=}2$ SCA will be given by
\begin{align}
\sT(z) &= \tfrac{1}{2k}\Omega^{\alpha\beta}(J_\alpha J_\beta) +
\tfrac{1}{2k}(\Omega^{\alpha\beta} + \tfrac{1}{k}{f^{\alpha}}_{i
\gamma} f^{i\gamma\beta}) (\d\Psi_\alpha\Psi_\beta)\notag\\
&\quad + \tfrac{1}{2k^2} f^{i\alpha\beta}(J_i\Psi_\alpha \Psi_\beta) -
\tfrac{1}{8k^3}{f_i}^{\gamma\delta} f^{i\alpha\beta} (\Psi_\alpha
\Psi_\beta \Psi_\gamma\Psi_\delta)~,\label{eq:ktcb1}\\
\sG_0(z) &= \tfrac{1}{k}\Omega^{\alpha\beta}(J_\alpha\Psi_\beta) -
\tfrac{1}{6k^2} f^{\alpha\beta\gamma} (\Psi_\alpha \Psi_\beta
\Psi_\gamma)~, \label{eq:ktcb2}\\
\sG_1(z) &= \tfrac{1}{k}A^{\alpha\beta}(J_\alpha\Psi_\beta) +
\tfrac{1}{6k^2} B^{\alpha\beta\gamma} (\Psi_\alpha \Psi_\beta
\Psi_\gamma)~,\label{eq:ktcb3}\\
\sJ(z) &= \tfrac{1}{2ik}A^{\alpha\beta}(\Psi_\alpha \Psi_\beta) -
\tfrac{1}{2ik}A^{\gamma\delta} {f_{\gamma\delta}}^c
I_c~,\label{eq:ktcb4}
\end{align}
where $J_a \equiv I_a - \tfrac{1}{2k}\Omega^{bd}{f_{ab}}^c
(\Psi_c\Psi_d)$ and $B^{\mu\nu\rho} = A^{\mu\alpha}A^{\nu\beta}
A^{\rho\gamma} f_{\alpha\beta\gamma}~$.  Notice that although so far
we have only considered the holomorphic sector, we have a similar
structure for the antiholomorphic sector as well.  In other words, we
have a $(2,2)$ SCFT.

Finally, we recall that one can describe the complex structure on
$\gg/\gh$ in an alternative way, by introducing the projection
operators $(P^{\pm})^\alpha_\beta = \half (\delta^\alpha_\beta \pm
\tfrac{1}{i}A^\alpha_\beta)$, which allow us to split the complexified
tangent space $\gt \cong (\gg/\gh)^\C$ into subspaces $\gt_+$ and
$\gt_-$ defined as the image of the projectors $P^+$ and $P^-$
respectively.  Introducing bases $\{X_{\alpha}^{\pm} =
(P^{\pm})_\alpha^\beta X_{\beta}\}$ for $\gt_\pm$ respectively, one
can then show that $\gt_\pm$ are lagrangian, and that $\gt$ admits a
decomposition $\gt = \gt_+ \oplus \gt_-$ into subspaces which close 
under the Lie brackets: $[\gt_\pm,\gt_\pm] \subset \gt_\pm$ (which
re-states the fact that the complex structure on $G/H$ is integrable).

\section{Boundary conditions for $N{=}2$ SCFTs}

D-branes can be been studied in a variety of ways: by using the
techniques of perturbative string theory, they can be described either
in terms of boundary conditions of open strings or as {\em boundary
states\/} in the closed string sector.  The concept of a boundary
state, which describes how closed strings are emitted or absorbed on
the D-brane worldvolume, allows us to perform perturbative string
computations in the presence of a D-brane and is thus instrumental in
considering D-branes in type II string theories. 

The guiding principle behind constructing a boundary state is conformal
invariance.  In open string theories one has to impose constraints on
the boundary conditions such that the (super)conformal symmetry is not
broken.  Then the boundary can be thought of as a closed string state
where the left-- and right--moving (super)conformal structures are
related in a {\em consistent\/} way.  Consistency means in this
context that the holomorphic SCFT is set equal to the antiholomorphic
SCFT, up to an automorphism of the $N{=}2$ SCA.  In other words
\begin{equation*}
\sA(z) = \tau(\bar\sA(\bar z))~,
\end{equation*}
where $\sA$ is a generic $N{=}2$ generator and $\tau$ is an
arbitrary element of the automorphism group of the $N{=}2$ SCA, that
is $O(2)$.  This ensures that the D-brane configuration preserves one
set of the $N{=}2$ SCA.

Further we require that our boundary state will possess {\em local\/}
$N{=}1$ worldsheet supersymmetry.  This means that the automorphism
defining the boundary conditions has to fix the $N{=}1$ subalgebra
generated by $(\sT,\sG_0)$, which leaves us with a $\Z_2 \times \Z_2$
group of transformations.

Finally we want the D-brane configuration to describe a BPS state, and
hence to preserve half of the spacetime supersymmetry.  This forces us
to extend the boundary conditions to the spectral flow operator.  In
the case of an $N{=}2$ SCFT spacetime supersymmetry is directly
related to the $U(1)$ current \cite{LVW}.  Indeed if one considers the
bosonisation of the $U(1)$ generator
\begin{equation}
\sJ(z) = i\d\phi\label{eq:u1c}
\end{equation}
then the spectral flow operator will be given by
\begin{equation}
X(z) = e^{i\phi}~.\label{eq:spf}
\end{equation}
This leaves us with the following sets of boundary conditions:

\begin{itemize}
{\bf \item[(i)] A-type boundary conditions}
\begin{equation}
\sT = \bar{\sT}~,\qquad \sG_0 = \pm \bar{\sG_0}~,\label{eq:A1}
\end{equation}
\begin{equation}
\sJ = - \bar{\sJ}~,\qquad \sG_1 = \mp \bar{\sG_1}~,\label{eq:A2}
\end{equation}
\begin{equation}
e^{\pm i\phi} = e^{\pm i\theta}e^{\mp i\bar{\phi}}~;\label{eq:A3} 
\end{equation}
{\bf \item[(ii)]B-type boundary conditions}
\begin{equation}
\sT = \bar{\sT}~,\qquad \sG_0 = \pm \bar{\sG_0}~,\label{eq:B1}
\end{equation}
\begin{equation}
\sJ = \bar{\sJ}~,\qquad \sG_1 = \pm \bar{\sG_1}~,\label{eq:B2}
\end{equation}
\begin{equation}
e^{\pm i\phi} = e^{\pm i\theta}e^{\pm i\bar{\phi}}~.\label{eq:B3}
\end{equation}
\end{itemize}

These boundary conditions have been written down previously in
\cite{OOY}, in the context of Calabi-Yau compactifications.  However 
they are clearly valid for a generic $N{=}2$ theory, independent of
any particular model.  In order to find solutions for them we will
have to specify a particular class of SCFTs---in our case the
Kazama--Suzuki model.  

\section{Boundary conditions for Kazama--Suzuki models}

The first step in solving our problem is finding a consistent set of
boundary conditions for the current algebra
\eqref{eq:affg1}-\eqref{eq:affg3} in terms of which the Kazama--Suzuki
model is defined, such that one of the two possible types of boundary
states will be realised.  (Alternatively one could argue that the
boundary conditions for the $N{=}2$ SCA are not restrictive enough to
determine the allowed configurations uniquely.)  This approach to
constructing boundary states, which goes back to \cite{I}, was used
recently in \cite{KO} for the study of bosonic D-branes on group
manifolds.  

Thus we require that the boundary conditions satisfied by the bosonic
and fermionic currents   
\begin{equation*}
I_a(z) = {R_a}^b {\bar I}_b(\bar z)~,\qquad 
\Psi_a(z) = {S_a}^b {\bar\Psi}_b(\bar z)~,
\end{equation*}
preserve the $N{=}1$ affine algebra \eqref{eq:affg1}-\eqref{eq:affg3},
which imposes the following conditions on the matrices $R$ and $S$:  
\begin{align}
{R_a}^c{R_b}^d\Omega_{cd} &= \Omega_{ab}~,\qquad
{R_a}^d{R_b}^e{f_{de}}^f = {f_{ab}}^c{R_c}^f~,\label{eq:condr}\\
{S_a}^c{S_b}^d\Omega_{cd} &= \Omega_{ab}~,\qquad 
{R_a}^d{S_b}^e{f_{de}}^f = {f_{ab}}^c{S_c}^f~.\label{eq:conds}
\end{align}

This immediately implies that the $N{=}1$ Sugawara energy-momentum
tensor $\sT_{\gg}(z) = \tfrac{1}{2k}\Omega^{ab}(J_a J_b) + \tfrac{1}{2k}
\Omega^{ab} (\d\Psi_a\Psi_b)$ constructed from $\widehat\gg_{N{=}1}$
satisfies  
\begin{equation}
\sT_{\gg} = \bar\sT_{\gg}~,\label{eq:bctg}
\end{equation}
at the boundary.  From the boundary condition on $\sG_0$ one deduces
that 
$\Omega^{\alpha\beta}{R_{\alpha}}^i{S_{\beta}}^a =
\Omega^{\alpha\beta}{R_{\alpha}}^a{S_{\beta}}^i = 0$ which in turn
implies that 
\begin{equation}
{R_{\alpha}}^i = {S_{\alpha}}^i = 0~.\label{eq:split}
\end{equation}
In other words $R$ and $S$ preserve $\gh$ and $\gg/\gh$.  Moreover,
from \eqref{eq:condr} and \eqref{eq:conds} and by using Schur's lemma
we deduce that
\begin{equation}
S = \pm R~,\label{eq:sr}
\end{equation}
as one would expect from supersymmetry.

On the other hand \eqref{eq:split} implies that the $N{=}1$ Sugawara
energy-momentum tensor $\sT_{\gh}(z) = \tfrac{1}{2k}\Omega^{ij}(J_i
J_j) + \tfrac{1}{2k} \Omega^{ij} (\d\Psi_i\Psi_j)$ constructed from
$\widehat\gh_{N{=}1}$ satisfies
\begin{equation*}
\sT_{\gh} = \bar\sT_{\gh}~,
\end{equation*}
which together with \eqref{eq:bctg} ensures that that boundary
condition corresponding to the Kazama--Suzuki energy-momentum tensor
$\sT$ is satisfied.  Similarly one can also show for the other $N{=}1$
Sugawara generators
\begin{equation*}
(\sG_0)_{\gg} = \pm(\bar\sG_0)_{\gg}~,\qquad 
(\sG_0)_{\gh} = \pm(\bar\sG_0)_{\gh}~,
\end{equation*}
which implies that the boundary conditions for the $\gg$ and $\gh$
$N{=}1$ SCAs are independently satisfied.

We now turn to the boundary conditions for $\sG_0$, which are the same
for both types A and B; they impose the following conditions:
\begin{align}
{R_{\alpha}}^{\gamma}{R_{\beta}}^{\delta}\Omega_{\gamma\delta} &=
\Omega_{\alpha\beta}~,\label{eq:rort}\\
{R_{\alpha}}^{\mu}{R_{\beta}}^{\nu}{R_{\gamma}}^{\rho}
f^{\alpha\beta\gamma} &= f^{\mu\nu\rho}~.\label{eq:rmor} 
\end{align}
Hence, according to the first relation, $R$ preserves the metric on
$\gg/\gh$.

The boundary conditions on $\sG_1$, on the other hand are slightly
different for the two types of boundary conditions, and relate $R$ to 
the complex structure on $G/H$:
\begin{align}
A^{\alpha\beta}{R_{\alpha}}^{\gamma}{R_{\beta}}^{\delta} &=
\mp A^{\gamma\delta}~,\label{eq:ar}\\ 
B^{\alpha\beta\gamma}{R_{\alpha}}^{\mu}{R_{\beta}}^{\nu}
{R_{\gamma}}^{\rho}  &= \mp B^{\mu\nu\rho}~,\label{eq:br} 
\end{align}
where the two signs refer to the A and B types of boundary conditions,
respectively.  It is easy to see that the second condition
\eqref{eq:br} follows from \eqref{eq:rmor} and \eqref{eq:ar}, hence it
does not yield any new information.  On the other hand, the first
condition \eqref{eq:ar} can be rephrased by saying that the complex
structure $A$ (anti)commutes with the matrix $R$.  If we now consider
the split of $\gt$ into the subspaces $\gt_+$ and $\gt_-$, then
\eqref{eq:ar} amounts to two sets of boundary conditions (one for each
case) for the generators of these subspaces: $X_{\alpha}^{\pm} =
{R_{\alpha}}^{\beta}{\bar X}_{\beta}^{\mp}$ for the A-type, and
$X_{\alpha}^{\pm} = {R_{\alpha}}^{\beta}{\bar X}_{\beta}^{\pm}$ for
the B-type boundary conditions where, in order to avoid confusion, we
denoted by $\{{\bar X}_{\alpha}\}$ the generators of $\gg/\gh$
corresponding to the antiholomorphic sector.  Symbolically we can
write this conditions as $\gt_{\pm} = R(\Bar\gt_{\mp})$ and $\gt_{\pm}
= R(\Bar\gt_{\pm})$, respectively.

One can check that the boundary conditions on the $U(1)$ current $\sJ$
do not provide any new conditions relating $R$ to the geometric data
of the target space.  Therefore from \eqref{eq:u1c} one can deduce
that $\phi = \mp \bar\phi + \theta$, where $\theta$ is a constant.  
However in order to properly analyse the boundary conditions
\eqref{eq:A3} and \eqref{eq:B3}, and in particular to fix $\theta$,
one would need an explicit expression of the spectral flow
\eqref{eq:spf} in terms of the affine currents (which are $N{=}1$
primaries) and the geometric data on the coset space. 

\section{Supersymmetric cycles}

We will study here the geometry of the supersymmetric cycles in
Kazama--Suzuki models.  By a configuration in which a D-brane wraps
around one such cycle, we mean one in which we identify the
worldvolume of the D-brane with the cycle or, more precisely, we
identify the tangent and normal directions to the cycle with the
tangent and normal directions to the worldvolume of the D-brane,
respectively.\footnote{Strictly speaking we mean the component of the
worldvolume of the D-brane in the internal theory defined by the
Kazama--Suzuki model.  From the point of view of an observer in the
spacetime, such a D-brane configuration could present itself
point-like, if the D-brane has no tangent directions in the spacetime,
or extended, if it does.  For euclidean D-branes, it could even
present itself as an instanton.  Our discussion clearly does not
depend on these details.}  We will restrict ourselves to the case
without mixed boundary conditions.  In this case the matrix $R$ is
diagonalisable, with the eigenvectors corresponding to the $(\pm
1)$-eigenvalues being identified with the Neumann and Dirichlet
boundary conditions on the bosonic currents, respectively.

It is easy to see that the Neumann directions are indeed tangent to
submanifolds of $G/H$; that is, to the D-branes.  We see this first at
a point in $G/H$ (the identity coset), where we can identify the
tangent space with $\gg/\gh$.  The Neumann directions correspond to
the $(+1)$-eigenvalues of the matrix $R$; let us call them
$(\gg/\gh)_+$.  We would like to interpret $(\gg/\gh)_+$ as the
tangent space of a submanifold of $G/H$ at that point.  This would
follow from the Frobenius integrability theorem if we could show that
vector fields tangent to the Neumann directions close under the Lie
bracket.  Let us prove this.  Under the action of $R$, $\gg$ also
decomposes as $\gg = \gg_+ \oplus \gg_-$, where $\gg_+$ is a
subalgebra.  Because $R$ is block diagonal, it follows that
$(\gg/\gh)_+$ is the projection of $\gg_+$ onto $\gg/\gh$.  The
projection $\gg \to \gg/\gh$ extends to a Lie algebra homomorphism
from $\gg$ to the vector fields on $G/H$. The image of this map are
the fundamental (or Killing) vectors generating infinitesimally the
$G$ action.  Because $\gg_+$ is a subalgebra, its image closes under
Lie brackets.  As a matter of fact, Frobenius's theorem tells us more:
it guarantees that every point in $G/H$ is contained in a unique
submanifold whose tangent vectors are the Killing vectors in $\gg_+$;
that is, $G/H$ is foliated by its D-branes.

This raises an interesting question; namely, whether this foliation
of the coset space by its D-branes bears any similarity to the
foliation of a Calabi-Yau space by its special lagrangian tori
\cite{SYZ,M}.  One could go even further and speculate about a
possible generalisation of the mirror conjecture of \cite{SYZ} in the
context of coset spaces.  The answer to these questions is beyond our
reach at the present moment, largely due to our insufficient
understanding of the geometry of coset spaces, in particular of their
calibrated submanifolds (as we will see in a moment).  However we
should point out that the origin of the two foliations is slightly
different: in the case of $G/H$ it is a consequence of the homogeneity
of the coset manifold, by contrast to Calabi-Yau spaces which are far
from being homogeneous.

Let us now consider an $n$-cycle $\gamma$ in our target space.  We
want to study the configurations corresponding to D-branes wrapping
around this cycle.  Locally we can choose coordinates (that is, a
basis for $\gg/\gh$) such that $\alpha = 1,...,n$ corresponds to the
directions tangent to $\gamma$, whereas $\mu = n+1,...,\dim G/H$
corresponds to the directions normal to the cycle.  We then have two
possible situations depending on the type of boundary state to which
the D-brane gives rise, when wrapped around $\gamma$.

\subsection*{A-type cycles}  

The D-brane wrapped around $\gamma$ gives rise to a boundary state
with type A boundary conditions.  Then the condition \eqref{eq:rort},
coming from the boundary condition satisfied by $\sG_0$, implies that
the metric $\Omega$ on $\gg/\gh$ has a block diagonal form with
respect to $\gamma$, that is $\Omega_{\alpha\mu} = 0$, which we can
write symbolically as follows:
\begin{equation}
\Omega = \Omega_{\gamma} + \Omega_{\gamma^{\perp}}~.\label{eq:metsp}
\end{equation} 
In other words the metric on our coset space $G/H$ naturally induces a
metric $\Omega_{\gamma}$ on the cycle.

The boundary condition on $\sG_1$, which for the A-type cycles is
given by \eqref{eq:ar} with the minus sign, implies that the 2-form 
$A$ defined by
\begin{equation*}
A(X_{\alpha},X_{\beta}) = A_{\alpha\beta}
\end{equation*}
vanishes when restricted to either $\gamma$ or $\gamma^{\perp}$.
Since $A$ is nondegenerate, it follows that the dimension of the
cycle has to be equal to half the dimension of $G/H$, that is
\begin{equation*}
n = \half~\dim G/H~.
\end{equation*}
The vanishing of $A$ means that the complex structure maps vectors
tangent to $\gamma$ to vectors tangent to $\gamma^\perp$; hence the
tangent space to $\gamma$ contains no complex lines.  In other words,
it is a totally real subspace of $\gg/\gh$.  By analogy with
symplectic geometry we will say that $\gamma$ is a lagrangian
submanifold; but it is important to emphasise that $A$ need not be
closed.  Indeed, using the integrability condition for the complex
structure one can show that
\begin{equation*}
dA (X_{\alpha},X_{\beta},X_{\gamma}) = - B_{\alpha\beta\gamma}~.
\end{equation*}
Hence $dA=0$ if and only if $B_{\alpha\beta\gamma} = 0$ which, given
that $A$ is nondegenerate, is equivalent to $f_{\alpha\beta\gamma} =
0$.  We therefore conclude that the 2-form $A$ is a K\"ahler form on
$G/H$ if and only if $G/H$ is a hermitian symmetric space (HSS).
This does not mean that $G/H$ non-symmetric might not possess a
$G$-invariant K\"ahler structure.  Indeed, a theorem of Borel
\cite{B} guarantees that this is the case for $G$ and $H$ compact,
provided $H$ is the centraliser of a torus.

\subsection*{B-type cycles}

We now consider a D-brane that wraps on a cycle $\gamma$ giving rise
to a boundary state of type B.  Since $\sG_0$ satisfies the same
boundary conditions as before we are lead to the same conclusion 
\eqref{eq:metsp} about the metric on the coset space.  On the other 
hand, the boundary conditions \eqref{eq:ar} (now with the plus sign)
will imply that $A$ has a block diagonal structure with respect to
$\gamma$, that is $A_{\alpha\mu}=0$, which we can write symbolically
as
\begin{equation*}
A = A_{\gamma} + A_{\gamma^{\perp}}~.
\end{equation*}
Since $A$ is nondegenerate it follows that both $A_{\gamma}$ and
$A_{\gamma^{\perp}}$ are nondegenerate as well.  Furthermore it
follows that the complex structure on $G/H$ induces an almost
complex structure on the cycle.  One can in fact show that
$A_{\gamma}$ satisfies the integrability condition.  Indeed, if we
recall \cite{FS1} that the integrability condition  for the complex
structure on $G/H$ (which is equivalent to the vanishing of its
Nijenhuis tensor) reads   
\begin{equation*}
f^{\mu\nu\rho} 
 - A^{\mu\alpha}A^{\nu\beta}{f_{\alpha\beta}}^{\rho} -
   A^{\nu\alpha}A^{\rho\beta}{f_{\alpha\beta}}^{\mu} - 
   A^{\rho\alpha}A^{\mu\beta}{f_{\alpha\beta}}^{\nu} = 0~,
\end{equation*}
we can easily deduce that also $A_{\gamma}$ satisfies such a condition
and  hence that it defines a complex structure on $\gamma$.  In other
words, $\gamma$ is a complex submanifold of $G/H$.  In
particular, its dimension must be even \footnote{ The nomenclature can
be slightly misleading here since the B-type boundary conditions seem
to describe even-dimensional D-branes which, on the other hand, appear
in type IIA theories.}
\begin{equation*}
n = 0,2,4,...,\dim G/H~.
\end{equation*}

\section{D-brane geometry}

In the previous section we have determined the geometric
characteristics of D-branes on Kazama--Suzuki models with target space
a homogeneous hermitian manifold $G/H$.  The two types of boundary
conditions give rise to two types of submanifolds:
\begin{itemize}
\item[(A)] lagrangian submanifolds, and
\item[(B)] complex submanifolds.
\end{itemize}
At first sight this looks very similar to the case of Calabi--Yau
manifolds, in which one gets special lagrangian and K\"ahler
submanifolds, respectively \cite{BBS,OOY}; these submanifolds are
distinguished in that they are examples of calibrated submanifolds.
In the case of coset models, the above submanifolds are not calibrated
-- in fact, they need not be minimal -- but they will be seen to
belong to a class of geometries which are not unrelated to calibrated
geometries. In order to explain this, it is necessary to digress
slightly into the subject of calibrated geometries \cite{HL}.

There is more than one way to formulate the geometric structure of a
manifold.  Perhaps the most common approach is to specify such a
structure through a distinguished family of tensor fields on the
manifold, e.g., the K\"ahler form; or by restricting the types of
coordinate transformations which are allowed between different charts,
e.g., $G$-structures.  A less common approach, but one which seems
particularly relevant for the study of D-branes, is to specify a
distinguished family of submanifolds.  One way of singling out a
family of submanifolds is via the method of calibrations.

A {\em calibration\/} in a riemannian manifold $M$ is a closed
$p$-form $\omega$ which, acting on any unit simple $p$-vector
$\xi$ in $\bigwedge^pT_xM$, satisfies $\omega(\xi)\leq 1$.
In other words, the restriction of $\omega$ to any $p$-plane is
less than or equal to the volume element.  A $p$-plane is said to be
{\em calibrated\/} by $\omega$ if $\omega$ coincides with the volume
element.  A $p$-dimensional submanifold $N$ is said to be {\em
calibrated\/} by $\omega$ if its tangent space $T_xN$ is calibrated by
$\omega_x$ for all $x$ in $N$.  Calibrated submanifolds minimise volume
in their homology class.  Indeed, if $N'$ is any other $p$-dimensional
submanifold homologous to $N$, then
\begin{equation*}
\vol N = \int_N \omega = \int_{N'} \omega \leq \vol N'~,
\end{equation*}
where the second equality uses Stokes' theorem and the fact that
$\omega$ is closed.

This concept has an interesting generalisation, which is called a
$\phi$-{\it geometry} \cite{HL}.  This is very similar to a
calibrated geometry, the only difference being that the form $\phi$
need not be closed.  This difference has as a consequence the fact
that the associated ``calibrated'' submanifolds are not globally
volume-minimising. 

On a compact manifold any closed form can be normalised so that it is
a calibration, but the computation of the normalisation is a very
difficult problem and so is the determination of the calibrated
submanifolds.   Luckily some types of manifolds come with ready-made
calibrations.  For example any complex submanifold of a K\"ahler
manifold is calibrated by the relevant power of the K\"ahler form;
also special lagrangian submanifolds of Calabi--Yau manifolds are
calibrated by the real part of the holomorphic volume form, whereas
associative, coassociative and Cayley submanifolds of manifolds with
exceptional holonomy are also calibrated by parallel forms naturally
appearing in those geometries. 

K\"ahler and special lagrangian submanifolds of a Calabi--Yau manifold
made their appearance in \cite{BBS,OOY} in the context of D-branes.
More precisely, it was shown that the worldvolume theory of a
euclidean D-brane is supersymmetric precisely when the submanifold on
which the D-brane wraps is either K\"ahler or special lagrangian.  The
volume minimisation property for these manifolds is then the geometric
restatement of the BPS condition.  This result is consistent with the
dynamics of euclidean D-branes, in that when the $B$-field and the
gauge field on the worldvolume of the D-brane are turned off, the
Dirac--Born--Infeld action is simply the volume of the D-brane.

In the case of Kazama--Suzuki models $G/H$, we have found that the
B-type cycles are complex submanifolds relative to the $G$-invariant
complex structure.  Wirtinger's inequality still applies and we
can conclude that the tangent planes to these submanifolds are
calibrated with respect to the normalised power of the nondegenerate
2-form $A$.  However, because $A$ is not closed in general, these
submanifolds fail to be minimal, and what we have is a {\it complex
geometry} defined by $A$.  

It is important to remark that this result does not conflict with the
requirement that these configurations be BPS.  We have seen that the
D-branes in Kazama--Suzuki models are not necessarily minimal
submanifolds, and this conclusion seems to run against our geometric
intuition, which leads us to expect that supersymmetric cycles be
calibrated or at least minimal.  However this expectation is born out
of the assumption that the mass of a BPS state obtained by a D-brane
wrapped on a supersymmetric cycle is given by its volume.  In the
cases where this has been shown to hold, the string couples only to
the background metric; but it is hard to argue convincingly that the
mass of the D-brane should not change in the presence of other
background fields: the $B$-field or the dilaton, for instance
(although see \cite{IMSS,SS}).  In the case of the Kazama--Suzuki
model, the string couples to the $B$-field (whose torsion is given by
the structure constants of $\gg$) and hence, strictly speaking, we
cannot conclude that a BPS state, while certainly having minimal mass,
also has minimal volume.

In the special case of $G/H$ a hermitian symmetric space the $B$-field
is absent (being essentially the torsion of the $H$-connection, which
vanishes in the symmetric case).  But in this case, $A$ is a closed
form, and the submanifolds calibrated by powers of $A$ are indeed
volume minimising. 

For the A-type cycles the situation is much less clear, due to the
fact that, lacking an explicit form for the spectral flow generator,
one is missing one extra condition.  One therefore expects that A-type
supersymmetric cycles will not just be lagrangian, but will be
somehow further constrained.  With the Calabi--Yau example in mind,
one would hope that a condition akin to the special lagrangian
condition should hold.


The problem of determining whether a homogeneous hermitian space $G/H$
possesses both A-type and B-type cycles is not an easy one.  B-type
cycles certainly always exist: $G/H$ itself is a complex submanifold
albeit not a proper one.  The situation for A-type cycles is not as
clear.  First of all it is easy to see that there are hermitian
symmetric spaces which do not possess nontrivial lagrangian
submanifolds; e.g., the 2-sphere $S^2 \cong SU(2)/U(1)$.  The
two-sphere is a K\"ahler manifold which possesses zero- and
two-dimensional complex submanifolds and one-dimensional lagrangian
submanifolds (isomorphic to $S^1$).  However, whereas the complex
submanifolds are indeed nontrivial calibrated cycles, $S^1$ is not
calibrated and in fact it is a trivial cycle.  This is symptomatic of
a more general situation: For compact homogeneous K\"ahler spaces the
odd cohomology groups are trivial \cite{B}, and this implies that for
coset spaces of dimension $4k+2$ there can be no nontrivial
middle-dimensional cycles.  On the other hand, when $\dim G/H = 4k$,
both types of cycles are even-dimensional; however finding nontrivial
lagrangian cycles for a given coset space is a difficult task.  (In
fact even for Calabi-Yau spaces, which have been extensively studied,
very few explicit examples of special lagrangian submanifolds are
known \cite{M}.)  Nevertheless they do exist.  For example, the coset
$SU(2+2k)/S(U(2)\times U(2k))$, which is an $8k$-dimensional hermitian
symmetric space of type $AI\!I\!I$ in Helgason's nomenclature, does
possess a minimal lagrangian submanifold.  To see this notice that
$SU(2+2k)/S(U(2)\times U(2k)) \cong G_{2,2k+2}(\C)$, the complex
grassmannian of complex 2-planes in $\C^{2k+2}$.  Hence the
grassmannian $G_{2,2k+2}(\R)$ of real 2-planes in $\R^{2k+2}$ is a
lagrangian submanifold which as shown in \cite{V} is globally minimal.
Notice that $G_{2,2k+2}(\C)$ also has K\"ahler submanifolds.  Indeed,
it is a compact homogeneous K\"ahler manifold and it is shown in
\cite{B} that any such manifold has a cellular decomposition in terms
of analytic cycles.  Hence for any $k$, the Kazama--Suzuki model with
target $G_{2,2k+2}(\C)$ possesses both types of D-branes.

Let us conclude with a remark.  We have obtained that the
supersymmetric cycles of Kazama--Suzuki models are given by a
distinguished set of submanifolds, which are ``locally calibrated'',
yet not globally volume-minimising.  We have also argued that this
does not conflict with the BPS requirement coming from the DBI action.
In order to make clearer the connection between the (geometric)
analysis of supersymmetry and the one of dynamics, one would need
a better understanding of the geometric content of the DBI action.
One can easily analyse in the limit in which the $B$-field vanishes.
This limit is to be taken in the In\"on\"u--Wigner sense \cite{IW}, as
the level is sent to infinity.  Looking at how the level enters in the
expression for the energy-momentum tensor, we see that this limit 
corresponds to the large volume limit in which, since the model is now
given by free bosons on a circle, the geometry of the target space
becomes toroidal.  In this case, the supersymmetric cycles which we
have found are indeed trivially calibrated.  One can argue that the
BPS condition remains true since this should not depend on the moduli,
but we have no right to expect the same for the calibrated condition.

\section{Abelian T--duality}

In this section we will set up the study of abelian T--duality for the
D-brane configurations we have analysed so far.  For this purpose we
will make use of the fact that Kazama--Suzuki models can be realised
as gauged supersymmetric WZW models \cite{W,FS1}. 

Let us start by briefly reviewing T--duality symmetry for WZW models
(for details see \cite{K,AAL,GW}).  The structure of the duality group
here is slightly different from the one present in flat backgrounds.
First of all the WZW model is {\em self-dual\/} under an abelian
T--duality transformation.  Moreover T--duality symmetry is realised
at the level of the affine algebra as Weyl transformations acting on
the primary currents.  Thus the symmetry underlying T--duality is
the invariance under the affine Weyl group.

T--duality for coset models (gauged WZW models) is intimately related
to the T--duality symmetry of the original WZW models.  In general one
can consider two limiting cases, according to whether the gauged
subgroup $H$ is semisimple or abelian.  If $H$ is semisimple then the
duality symmetry of the coset is inherited from that of the original
WZW model:  any dual pair of coset models is obtained by gauging a
dual pair of WZW actions.  In our case, since we want $G/H$ to be
K\"ahler, we only consider subgroups such that
$\rank\gg=\rank\gh$.  The duality transformations of the coset
model are then obtained by considering those duality transformations
at the level of the original theory which act trivially at the level
of $H$.  (In other words we consider only those Weyl transformations
which act trivially on the $\gh$ currents.)  This implies that a
duality transformation on the Kazama--Suzuki model will act trivially
on the Cartan subalgebra of $\gh$ which, assuming a regular embedding
of $\gh$ in $\gg$, coincides with the one of $\gg$, thus leaving all
the currents invariant.  Hence we are only left with trivial
T--duality transformations.   

We will therefore consider coset theories where the gauged subgroup is
abelian, hence of the form $G/{U(1)^\ell}$, where $\ell = \rank\gg$.
In this case a dual pair of cosets consists in two theories obtained
by performing either a vector or an axial gauging of a $U(1)$.  The
original affine Weyl symmetry of the WZW model guarantees that this
axial--vector duality is an exact symmetry.

By a T--duality transformation we will understand a map, at the level
of the fields, which preserves the $(2,2)$ superconformal structure of
the model.  More precisely, we will consider a map that acts trivially
on the antiholomorphic sector of the theory and is given by an
automorphism of $N{=}2$ SCA (whose automorphism group is $O(2)$) on
the holomorphic sector.  On the other hand, one can argue that if the
original and the dual theories are to describe the same physics then
they should couple in the same way to the gauge fields, and therefore
the T--duality transformation should fix the $N{=}1$ subalgebra
generated by $(\sT,\sG_0)$.  This leaves us with a $\Z_2 \times \Z_2$
group of transformations, that basically distinguishes two types of
T--duality, which we will denote by $T_A$ and $T_B$, respectively:
\begin{xalignat*}{3}
\sT &\to \sT~,\qquad& \sT &\to \sT~,\\
\sG_0 &\to \pm \sG_0~, \qquad& \sG_0 &\to \pm\sG_0~,\\ 
\sG_1 &\to \mp \sG_1~, \qquad& \sG_1 &\to \pm\sG_1~,\\ 
\sJ &\to - \sJ~,\qquad& \sJ &\to \sJ~.
\end{xalignat*}
$T_A$--duality is what one usually calls mirror symmetry \cite{M,SYZ},
but in our context we will treat both types of T--dualities on an
equal footing.   

At the level of the current algebra this T--duality map acts as a Weyl
transformation on the holomorphic currents.  In our basis this will be
described by a linear transformation  
\begin{equation*}
I'_a = {T_a}^b I_b~,\qquad \Psi'_a = {U_a}^b \Psi_b~,
\end{equation*}
while leaving invariant the currents in the antiholomorphic sector 
\begin{equation*}
{\bar I}'_a = {\bar I}_a~,\qquad {\bar\Psi}'_a = {\bar\Psi}_a~.
\end{equation*}
In this way the antiholomorphic SCA will be trivially preserved; for
the holomorphic sector we impose that the T--duality map at the level
of the currents will induce a $T_A$ or $T_B$ transformation of the
corresponding SCA. 

The analysis of the linear maps $T$ and $U$ is completely analogous to
the one of the matrices $R$ and $S$ describing the boundary
conditions. Since the essential symmetry here is the Weyl symmetry of
the ungauged WZW model, we start by defining the T--duality
transformation on the $\gg$ currents and impose that it preserve the
$N{=}1$ algebra \eqref{eq:affg1}-\eqref{eq:affg3}.  From this we
obtain that $U=\pm T$, that $T$ preserves the metric $\Omega$ and
gives rise to a Lie algebra automorphism on $\gg$.  From the
transformation law of $\sG_0$ we also deduce that $T$ has a block
diagonal structure with respect to $\gh$.  

Since T--duality is realised as Weyl transformations on the currents
it follows that the orthogonal matrix $T$ can be taken to be symmetric
and such that $T^2 = \1$.  In the previous sections we have seen that
the requirement for a boundary state to have a geometric
interpretation as a D-brane implies that the matrix $R$ defining that
boundary state is symmetric and squares to the identity matrix.  If we
now require that also the T--dual configuration defined by the
matrices $\tilde R$ and $\tilde S$
\begin{equation*}
I'_a(z) = {\tilde R_a}^b {\bar I}'_b(\bar z)~,\qquad 
\Psi'_a(z) = {\tilde S_a}^b {\bar\Psi}'_b(\bar z)~,
\end{equation*}
have such a geometric interpretation, then we can deduce that the 
T--duality transformation commutes with the original D-brane
configuration, $RT = TR$.  In other words $R$ and $T$ are
simultaneously diagonalisable. 

We now analyse the two types of duality transformations separately.
Notice first that $T_A$ maps from one type of boundary condition to
the other.  Furthermore, from the action of $T_A$ on $\bar\sG_1$ we
obtain 
\begin{equation}
A^{\alpha\beta}{T_{\alpha}}^{\gamma}{T_{\beta}}^{\delta} =
- A^{\gamma\delta}~,\label{eq:aTA}
\end{equation}
or, in other words, $T$ and $A$ anticommute.  From this one can deduce
that the order of the $T$ transformation, by which we mean the number
of $(-1)$ eigenvalues, is 
\begin{equation*}
\ord T_A = \half \dim G/H~.
\end{equation*}
We therefore consider two different cases according to the dimension
of the coset space:  
\begin{itemize}
\item[(i)] $\dim G/H = 4k$;

In this case $\ord T_A = 2k$, and hence such a T--duality
transformation will map within the same Type II theory.  For example,
if we start with a configuration corresponding to a D-brane which
wraps around a $2k$-dimensional A-type cycle $\gamma$, and we perform
a $T_A$--duality transformation, we end up with a configuration where
a D-brane wraps around an even dimensional B-type cycle $\gamma'$.
On the other hand, the $T_A$--dual of an even dimensional cycle
(corresponding to a holomorphic submanifold) will be a
$2k$-dimensional cycle corresponding (in the HSS case) to a lagrangian
submanifold.  This is summarised in Table~\ref{tab:TA2k}.

\begin{table}[h!]
\renewcommand{\arraystretch}{1.1}
\begin{tabular}{|c|c|c|c|}
\hline
IIA & IIB & A-type & B-type \\
    &     & $\dim\gamma= 2k$ & $\dim\gamma= 0,2,\ldots,4k$\\
\hline
IIA & IIB & B-type &     A-type\\
    &     & $\dim\gamma'= 0,2,\ldots,4k$ &$\dim\gamma'= 2k$ \\
\hline
\end{tabular}
\vspace{8pt}
\caption{Effect of a $T_A$ duality transformation of order $2k$: the
top row gets mapped to the bottom one.\label{tab:TA2k}}
\end{table}

\item[(ii)] $\dim G/H = 4k+2$;

In this case $\ord T_A = 2k+1$, and therefore $T_A$ will map from Type
IIA to Type IIB and vice versa.  The $T_A$--dual of a
($2k+1$)-dimensional A-type cycle will be an even dimensional B-type
cycle, and vice versa.  This is summarised in Table~\ref{tab:TA2k+1}.
\end{itemize}

\begin{table}[h!]
\renewcommand{\arraystretch}{1.1}
\begin{tabular}{|c|c|c|c|}
\hline
 IIA & IIB & A-type & B-type \\
     &     & $\dim\gamma= 2k+1$ & $\dim\gamma= 0,2,\ldots,4k+2$\\
\hline
 IIB & IIA & B-type &     A-type\\
     &     & $\dim\gamma'= 0,2,\ldots,4k+2$ & $\dim\gamma'= 2k+1$\\
\hline
\end{tabular}
\vspace{8pt}
\caption{Effect of a $T_A$ duality transformation of order $2k+1$: the
top row gets mapped to the bottom one.\label{tab:TA2k+1}}
\end{table}

Finally, we turn now to the $T_B$--duality map.  This maps between
boundary states of the same type.  Moreover, if we consider its
action on $\sG_1$ which implies
\begin{equation}
A^{\alpha\beta}{T_{\alpha}}^{\gamma}{T_{\beta}}^{\delta} =
 A^{\gamma\delta}~,\label{eq:aTB}
\end{equation}
we obtain for the order of $T_B$
\begin{equation*}
\ord T_B = 0,2,4,..., \dim G/H~,
\end{equation*}
from where we can conclude that it also maps within the same Type II
theory.  If we start with a configuration corresponding to a D-brane
wrapping around an even dimensional B-type cycle $\gamma$, and we
perform a $T_B$--duality transformation, we end up with a
configuration where a D-brane wraps around another even dimensional
B-type cycle $\gamma'$.  The other possibility is to start with an
A-type cycle $\gamma$ of dimension $\half\dim G/H$: in this case the
$T_B$--dual cycle $\gamma'$ will be will be of the same  
dimension, which means that $T_B$ has to act nontrivially in an equal
number of Neumann and Dirichlet dimensions.  The effect of a $T_B$
duality transformation is summarised in Table~\ref{tab:TB}.

\begin{table}[h!]
\renewcommand{\arraystretch}{1.1}
\begin{tabular}{|c|c|c|c|}
\hline
IIA & IIB & A-type & B-type \\
    &     & $\dim\gamma= \half\dim G/H$ & $\dim\gamma=
                                       0,2,\ldots,\dim G/H$\\ 
\hline
IIA & IIB & A-type &     B-type\\
    &     & $\dim\gamma'= \half\dim G/H$ & $\dim\gamma'=
                                      0,2,\ldots,\dim G/H$\\
\hline
\end{tabular}
\vspace{8pt}
\caption{Effect of a $T_B$ duality transformation: the top row gets
mapped to the bottom one.\label{tab:TB}}
\end{table}

One has to stress that all these T--dual pairs are ``subject to
availability''.  By this we mean that the existence of the
corresponding cycles $\gamma$ in the coset manifold is not guaranteed
a priori and has to be looked at case by case.

\section*{Acknowledgements}
It is a pleasure to thank Ralph Blumenhagen, Michael Flohr, Beatriz
Gato-Rivera, Christoph Schweigert and Arkady Tseytlin for valuable
discussions.  In addition I would like to thank Jos\'e
Figueroa-O'Farrill for getting me interested in this topic and for a 
critical reading of the paper.  I would also like to thank the referee
for useful criticism of an earlier version of this paper.  Finally, I
would like to acknowledge the support of the Isaac Newton Institute
for Mathematical Sciences, Cambridge, where this work was started.


\begin{thebibliography}{10}

\bibitem{BBS}
K.~Becker, M.~Becker, and A.~Strominger, ``Fivebranes, membranes and
  non-perturbative string theory,'' {\em Nuc. Phys.} {\bf 456} (1995) 130. {\tt
  hep-th/9509175}.

\bibitem{OOY}
H.~Ooguri, Y.~Oz, and Z.~Yin, ``D-branes on {C}alabi--{Y}au spaces and their
  mirrors,'' {\em Nuc. Phys.} {\bf B477} (1996) 407--430. {\tt hep-th/9606112}.

\bibitem{BBMOOY}
K.~Becker, M.~Becker, D.~R. Morrison, H.~Ooguri, Y.~Oz, and Z.~Yin,
  ``Supersymmetric cycles in exceptional holonomy manifolds and {C}alabi--{Y}au
  4-folds,'' {\em Nuc. Phys.} {\bf B480} (1996) 225. {\tt hep-th/9608116}.

\bibitem{HL}
F.~R. Harvey and H.~B. Lawson, ``Calibrated geometries,'' {\em Acta Math.} {\bf
  148} (1982) 47--157.

\bibitem{H}
F.~R. Harvey, {\em Spinors and calibrations}.
\newblock Academic Press, 1990.

\bibitem{BSV}
M.~Bershadsky, V.~Sadov, and C.~Vafa, ``D-branes and topological field
  theory,'' {\em Nuc. Phys.} {\bf B463} (1996) 420. {\tt hep-th/9511222}.

\bibitem{KO}
M.~Kato and T.~Okada, ``D-branes on group manifolds,'' {\em Nuc. Phys.} {\bf
  B499} (1997) 583. {\tt hep-th/9612148}.

\bibitem{BT}
M.~Blau and G.~Thompson, ``Aspects of ${N}_{T} \geq 2$ topological gauge
  theories and {D}-branes,'' {\em Nuc. Phys.} {\bf B492} (1997) 545--590. {\tt
  hep-th/9612143}.

\bibitem{BT2}
M.~Blau and G.~Thompson, ``Euclidean {SYM} theories by time reduction and
  special holonomy manifolds,'' {\em Phys. Lett.} {\bf B415} (1997) 242. {\tt
  hep-th/9706225}.

\bibitem{AFOS}
B.~S. Acharya, J.~M. Figueroa-O'Farrill, M.~O'Loughlin, and B.~Spence,
  ``Euclidean {D}-branes and higher-dimensional gauge theory,'' {\em Nuc.
  Phys.} {\bf B514} (1998) 583--602. {\tt hep-th/9707118}.

\bibitem{KS1}
Y.~Kazama and H.~Suzuki, ``Characterization of {$N{=}2$} superconformal models
  generated by coset space method,'' {\em Phys. Lett.} {\bf B216} (1989) 112.

\bibitem{KS2}
Y.~Kazama and H.~Suzuki, ``New {$N{=}2$} superconformal field theories and
  superstring compactification,'' {\em Nuc. Phys.} {\bf B321} (1989) 232.

\bibitem{S}
C.~Schweigert, ``On the classification of {$N{=}2$} superconformal coset
  theories,'' {\em Comm. Math. Phys.} {\bf 149} (1992) 425.

\bibitem{FS1}
J.~M. Figueroa-O'Farrill and S.~Stanciu, ``Nonreductive {WZW} models and their
  {CFT}s. {II}: {$N{=}1$} and {$N{=}2$} cosets,'' {\em Nuc. Phys.} {\bf B484}
  (1997) 583--608. {\tt hep-th/9605111}.

\bibitem{LVW}
W.~Lerche, C.~Vafa, and N.~P. Warner, ``Chiral rings in {$N{=}2$}
  superconformal theories,'' {\em Nuc. Phys.} {\bf B324} (1989) 427.

\bibitem{I}
N.~Ishibashi, ``The boundary and crosscap states in conformal field theories,''
  {\em Mod. Phys. Lett.} {\bf A4} (1989) 251.

\bibitem{SYZ}
A.~Strominger, S.~T. Yau, and E.~Zaslow, ``Mirror symmetry is {T}--duality,''
  {\em Nuc. Phys.} {\bf B479} (1996) 243--259. {\tt hep-th/9606040}.

\bibitem{M}
D.~R. Morrison, ``Mirror symmetry and the type {II} string,'' {\em Nucl. Phys.
  Proc. Suppl.} {\bf 46} (1996) 146--155. {\tt hep-th/9512016}.

\bibitem{B}
A.~Borel, ``K\"ahlerian coset spaces of semi-simple {L}ie groups,'' {\em Proc.
  Nat. Acad. Sci. USA} {\bf 40} (1954) 1147--1151.

\bibitem{IMSS}
H.~Ishikawa, Y.~Matsuo, Y.~Sugawara, and K.~Sugiyama, ``{BPS} mass spectrum
  from {D}-brane action.'' {\tt hep-th/9605023}.

\bibitem{SS}
Y.~Sugawara and K.~Sugiyama, ``D-brane analyses for {BPS} mass spectra and
  {U}-duality.'' {\tt hep-th/9707205}.

\bibitem{V}
L.~H. Van, ``Minimal surfaces in homogeneous spaces,'' {\em Math. USSR
  Izvestiya} {\bf 32} (1989) 413--427.

\bibitem{IW}
E.~In\"on\"u and E.~P. Wigner, ``On the contraction of groups and their
  representations,'' {\em Proc. Nat. Acad. Sci. USA} {\bf 39} (1956) 510--524.

\bibitem{W}
E.~Witten, ``The {$N$}--matrix model and gauged {WZW} models,'' {\em Nuc.
  Phys.} {\bf B371} (1992) 191. {\tt hep-th/9606040}.

\bibitem{K}
E.~Kiritsis, ``Exact duality symmetries in {CFT} and string theory,'' {\em Nuc.
  Phys.} {\bf B405} (1993) 109--142. {\tt hep-th/9302033}.

\bibitem{AAL}
E.~Alvarez, L.~Alvarez-Gaum\'e, and Y.~Lozano, ``A canonical approach to
  duality transformations,'' {\em Phys. Lett.} {\bf B336} (1994) 183--189. {\tt
  hep-th/9406206}.

\bibitem{GW}
A.~Giveon and E.~Witten, ``Mirror symmetry as a gauge symmetry,'' {\em Phys.
  Lett.} {\bf B332} (1994) 44--50. {\tt hep-th/9404184}.

\end{thebibliography}

\providecommand{\href}[2]{#2}\begingroup\raggedright\endgroup

\end{document}